\documentclass{article}
\usepackage{amsfonts,amssymb, amsmath}
\usepackage[english]{babel}

\def\bq{ \begin{equation}}
\def\eq{ \end{equation}}
\def\ben{ \begin{eqnarray}}
\def\en{ \end{eqnarray}}

\newtheorem{prop}{Proposition}

\begin{document}

%%%%%%%%%%%% TITLE %%%%%%%%%%%%%%

\title{Rotations and integrability}

\author{ A.V.Tsiganov\\
\it\small Steklov Mathematical Institute of Russian Academy of Sciences\\
\it\small e--mail: andrey.tsiganov@gmail.com}

\date{}
\maketitle

\begin{abstract}
We discuss some families of integrable and superintegrable systems in $n$-dimensional Euclidean space which are invariant to $m\geq n-2$
rotations. The integrable invariant Hamiltonian $H=\sum p_i^2+V(q)$ commutes with $n-2$ integrals of motion $M_\alpha $ and an additional integral of motion $G$, which are polynomials of first and fourth-order in momenta, respectively.
 \end{abstract}

\section{Introduction}
\setcounter{equation}{0}
This note represents a continuation of the research \cite{ts24} undertaken with regard to the construction of integrable and superintegrable Hamiltonian systems that are invariant under the action of a subgroup of the rotation group $SO(n)$ of Euclidean space $\mathbb R^n$.  Equations of motion for a natural Hamiltonian system in $\mathbb R^n$ are
\[
\dot{q}_i=\frac{\partial H}{\partial p_i}\,,\qquad \dot{p}_i=-\frac{\partial H}{\partial q_i}\,,\qquad i=1..n\,,
\] 
where
\[
H=\sum p_i^2+V(q_1,\ldots,q_n)\,.
\]
We suppose that potential $V(q)$ is invariant to the action of group $G\subset SO(n)$ such that its regular orbits are of dimension $n-2$.
In addition to the requisite additions and corrections to \cite{ts24}, the primary focus is on the examination of diverse forms of invariant potentials $V$ associated with distinct realizations of symmetry group $G$.

In classical mechanics, a Hamiltonian system on $2n$-dimensional symplectic manifold is called integrable by Liouville, if there are $n$ functionally independent integrals of motion in the involution. Integrable by Liouville system  is called non-degenerate system
if the Kolmogorov condition \cite{kolm59, kolm26,kolm27} for the Hessian matrix
\[
\mbox{det}\left|\left|\frac{\partial^2 H}{\partial I_i\partial I_j}\right|\right|\neq 0
\]
holds almost everywhere. Trajectories of the completely integrable non-degenerate Hamiltonian system
are everywhere dense on almost all tori. Therefore any smooth first
integral $F(I,\varphi)$ of the system  is constant on all tori and hence any first
integral $F$ is a function of the action variables $I=(I_1,\ldots,I_n)$ only:
\[\frac{dF}{dt}= 0 \quad \Rightarrow\quad F = F(I_1,\ldots,I_n).\]
Here $I_1,\ldots,I_n$ and $\varphi_1,\ldots,\varphi_n$ are the action-angle variables in a neighborhood of an invariant toroidal domain.
Kolmogorov was the first to mention the existence of Hamiltonian systems with invariant tori of dimensions $k>n$ on $2n$-dimensional symplectic manifolds, see \cite{kolm27}, p. 326. The invariant tori of general integrable systems are not necessarily Lagrangian, isotropic or coisotropic with respect to the
invariant symplectic structure.

A superintegrable system is, roughly speaking, a system that allows more global first integrals than degrees of freedom, see the exact definition in \cite{bbm14,mil,resh} and references therein.These global integrals of motion  are well-defined functions on the whole phase space and, simultaneously, they are functions on the both local action and angle variables  \cite{ts08}. Thus, superintegrable systems are always degenerate in the Kolmogorov sense.

If degenerate by Kolmogorov systems are integrable in Abel's quadratures, then additional global integrals of motion can be obtained using classical addition theorems \cite{ts08a,ts09},  theorems on the integration of binomial differentials \cite{ts18}, etc. If we do not know how to reduce the equations of motion to Abel's quadratures, we can use either the brute force method  or some other methods to construct the additional integrals of motion \cite{bbm14,hiet87,mil}.  Below we  discuss construction of the non-degenerate and degenerate integrable systems in the framework of symplectic reduction theory.

Symplectic reduction theory concerns the removal of variables using symmetries and their associated conservation laws, see \cite{cb97,mw2001} and references therein. This construction appears in classical works by Euler, Lagrange, Hamilton, Jacobi, Routh, Riemann, Liouville, Lie and Poincar\'{e}. Their work aimed to eliminate variables associated with symmetries to simplify calculations in concrete examples. The modern theory of geometric mechanics allows a unified description of known results and several new ones to be derived. Let us add to the textbooks \cite{cb97,mw2001}  a few more recent papers on the reduction of systems with symmetries, namely \cite{bl01,brav,mar14, ort04,sh06}. 
It is also important to highlight a link between reduction and partial separation of variables, for which there is currently a thriving body of research \cite{temp24}.

In this note, we propose a unified description for some integrable and superintegrable Hamiltonian systems in $n$-dimensional Euclidean space $\mathbb E^n$ which were obtained in 1983-1986, see \cite{f83, d86,f86,d85,r86}. In 1983 Fordy and Kulish studied Newton's equations of motion
\[
 \ddot{q}^\alpha=\sum_{\beta,\gamma,\delta} \mathcal R^\alpha_{\beta,\gamma,-\delta}q^\beta q^\gamma q^\delta,\qquad \alpha,\beta,\gamma,\delta=1,\ldots, n\,,
 \]
 associated with the nonlinear matrix Schr\"{o}dinger equations and $N$-wave equations hierarchies \cite{f83}. Here $R^\alpha_{\beta,\gamma,-\delta}$ is a curvature tensor on one of the reductive homogeneous spaces and, hence, the corresponding potential 
\[
V(q)=\sum_{\alpha,\beta,\gamma,\delta} \mathcal R_{-\alpha,\beta,\gamma,-\delta}q^\alpha q^\beta q^\gamma q^\delta
\]
is invariant to rotations which form subgroups $SO(k)\times SO(m)$ of the rotation group $SO(n)$. For all these systems we know Lax pairs, classical $r$-matrices and a possible integrable generalisation \cite{f86, r86}, see also textbooks \cite{pe90,rs94,tf94,rs03}.

In 1985 a method for constructing $n$-dimensional integrable Hamiltonian systems from two-dimensional ones was proposed in \cite{d85}. 
This method involves the addition of variables using symmetries and can therefore be considered as an inversion of the symplectic reduction method. The resulting $n$-dimensional systems are invariant to the action of the subgroup $SO(n-1)$ of the rotation group $SO(n)$. At $n=3$ the Painlev\'{e} analysis of the corresponding systems with quartic potentials is discussed in \cite{d86}. 

We reproduce these known results and obtain new integrable systems by combining the method of invariants 
which seeks to parameterize quotient space by group invariant functions or invariant coordinates \cite{ mw2001,cb97}, and the direct methods for the search of the first integrals \cite{hiet87}.

\section{Integrable systems invariant to rotations}
Following the classical works of the 18th and 19th centuries, Cartesian coordinates are used to describe concrete integrable systems. This is because the connection between mechanical systems with symmetries and geometry is now evident \cite{mw2001}. 

Using an orthonormal basis of a Euclidean vector space $\mathbb E^n$ we identify the rotation group with the group of orthogonal matrices $SO(n)$. In this article rotation means rotational displacement or so-called rigid rotation, see all the necessary definitions in the Cayley kinematics  \cite{cayley}.

Rotationally invariant potentials $V(q)$ on Euclidean space $\mathbb E^n$ satisfy a system of equations
\bq\label{m-eq}
\mathcal L_{Y_\alpha}\,V(q)=0\,,\qquad \alpha=1,\ldots,m
\eq
where $\mathcal L$ is a Lie derivative along the Killing vector fields 
 \[
Y_\alpha=\sum c^{ij}_\alpha X_{ij}\qquad c_\alpha^{ij}\in \mathbb R\,,
\]
forming a subalgebra $S_m$ of the Lie algebra $ so(n)$ of the special orthogonal group $SO(n)$. Here
\[
X_{ij}=q_i\partial_j-q_j\partial_i\,,\qquad \partial_k=\frac{\partial}{\partial q_k}\,,
\]
are vector fields associated with the principal rotations of the $ij$ planes in $\mathbb E^n$. According to Noether's theorem, there are conservation laws
 \[\label{M-a}
 M_\alpha=\sum c_\alpha^{ij} J_{ij}\,, \qquad J_{ij}=q_ip_j-q_jp_i
 \]
in the involution 
\[
\{H,M_\alpha\}=0\,, \qquad \alpha=1,\ldots,m
\]
with the invariant Hamiltonian 
\bq\label{ham}
H=T+V=\sum_{i=1}^n p_i^2+V(q)\,.
\eq
Here $\{.,.\}$ is canonical Poisson bracket on a symplectic manifold $T^*\mathbb E^n$
\bq\label{poi}
\{q_i,q_j\}=0\,,\qquad \{p_i,p_j\}=0\,,\qquad \{q_i,p_j\}=\delta_{ij}\,.
\eq
The symmetry fields $Y_\alpha$, $\alpha=1,\ldots,m$ form a subalgebra $S_m$ of $so(n)$, which can be abelian or non-abelian. In both cases, we assume that these Killing vector fields generate $n-2$ independent integrals of motion in the involution for each other. 
Thus, we get an integrable system in the Liouville sense, if we find another independent integral of motion $G(q,p)$ in the involution to the other integrals of motion. 

For the search of the second invariant $G(q,p)$ we use a combination of a direct method \cite{hiet87} with a method of invariants \cite{mw2001}. The method of invariants searches for the parameterisation of quotient spaces by the group invariant functions. It has a rich history going back to Hilbert's invariant theory, see the book by Cushman and Bates \cite{cb97} for more details and references.

\subsection{Invariant variables}
The Hamiltonian $H=T+V$ (\ref{ham}) consists of the kinetic and potential parts which we consider as two invariant coordinates on a quotient space $T^*\mathbb E^n/S_m$ 
\[P_1=\sum_{i=1}^n p_i^2\qquad\mbox{and}\qquad Q_2=V(q)\,.\] 
Using canonical transformation 
\[
\varphi:\qquad (q,p)\to (p,-q)
\]
preserving both symplectic structure and angular momentum tensor $J$ with entries $J_{ij}=q_ip_j-q_jp_i$, we obtain second pair of invariant variables
\[
Q_1=\sum_{i=1}^n q_i^2\qquad\mbox{and}\qquad P_2=V(p)\,.
\] 
The Poisson brackets between invariant variables are also rotationally invariant variables. 

For instance, pair of "kinetic" variables $P_1$, $Q_1$ and the Poisson bracket between them
\[A=\{P_1,Q_1\}=-4\sum_{i=1}^n q_ip_i\]
are in the involution with any component $J_{ij}=q_ip_j-q_jp_i$ of the angular momentum tensor
\[\{Q_1,J_{ij}\}=0\,,\quad\{P_1,J_{ij}\}=0\,,\quad \{A,J_{ij}\}=0\,,\qquad i,j=1,\ldots,n\,.\]
The canonical Poisson brackets (\ref{poi}) between these functions 
\[
\{P_1, A\} =8P_1\,,\qquad \{Q_1, A\}=- 8Q_1\,,\qquad \{P_1,Q_1\}=A
\]
 Are equivalent to the Lie-Poisson brackets between elements of the algebra $so^*(3)$ up to the scaling. The corresponding Casimir function for these Lie-Poisson brackets is a sum of the squares of the components of the angular momentum tensor
\[
J^2\equiv\sum_{i>j}^n J_{ij}^2=P_1Q_1-\frac{1}{16}A^2\,.
\]
Using known relations between the classical angular momentum theory and the Gaudin magnets theory \cite{kuz92} we can identify variables $P_1, Q_1$ and $A$ with the components of the total spin of the corresponding Gaudin magnet. 

Second set of "potential" variables $Q_2=V(q)$ and $P_2=V(p)$ are polynomials of the $N$-ts order on variables $p_i$ and $q_i$. As a result, a quotient space $T^*\mathbb E^n/S_m$ is a semidirect product of the Lie-Poisson algebra $so^*(3)$ with the Poisson space having a polynomial bracket.

Any non-trivial linear solution of the equations (\ref{m-eq}) 
\[V(q)=\sum_{i=1}^n V_i q_i\,,\qquad V_i\in\mathbb R\,,\] 
can be reduced to solution invariant to the action of subgroup $SO(n-1)\subset SO(n)$
\[V(q)=q_n\]
using rotations and scaling. The corresponding quotient space is parameterized by five invariant coordinates 
\[Q_0= V(q)\,,\quad P_0= V(p)\,,\quad Q_1=\sum_{i=1}^n q_i^2\,,\quad P_1=\sum_{i=1}^n p_i^2\,,\quad A=-4\sum_{i=1}^n q_ip_i\] 
with the Poisson brackets defined by the non-homogeneous linear Poisson bivector
\[
\Pi=\left(
 \begin{array}{ccccc}
 0 & 1 & 0 & 2P_0 & -4Q_0 \\
 -1 & 0 & -2Q_0 & 0 & 4P_0 \\
 0 & 2Q_0 & 0 & -A & -8Q_1 \\
 -2P_0 & 0 & A & 0 & 8P_1 \\
 4Q_0 & -4P_0 & 8Q_1 & -8P_1 & 0 \\
 \end{array}
 \right)\,,
\]
and the Casimir function
\[
C=8AP_0Q_0 + 16P_0^2Q_1 + 16P_1Q_0^2 + A^2 - 16P_1Q_1\,,
\]
so that $\Pi\, dC=0$. We can use these invariant variables to construct generalizations of the H\'{e}non-Hailes, Holt and quartic oscillator systems \cite{d86,d85}. 

There are subgroups of $SO(n)$ for which the basis of solutions (\ref{m-eq}) consists of polynomials of higher order in $q$. For instance, for a pair of commuting rotations in $\mathbb E^4$ 
\[
Y_1=X_{12}=q_1\partial_2-q_2\partial_1\,,\qquad Y_2=X_{34}=q_3\partial_4-q_4\partial_3
\]
basis of the solutions consists of two polynomials of second-order
\[
Q_1=q_1^2+q_2^2+q_3^2+q_4^2\,,\qquad Q_2=q_1^2+q_2^2-q_3^2-q_4^2\,.
\]
For the so-called double isoclinic rotations in $\mathbb E^4$ 
\[
Y_1=X_{12}+X_{34}\,,\qquad Y_2=X_{13}+X_{24}
\]
basis of the solutions consists of polynomials
\[
Q_1=q_1^2+q_2^2+q_3^2+q_4^2\,,\qquad Q_2=q_1q_4 - q_2q_3\,.
\]
Classifications of the subgroups of $SO(n)$, subalgebras of $so(n)$ and their representations by solutions $V(q)$ of equations (\ref{m-eq}) are closely related to the Cartan classification of the homogeneous space \cite{f83}. 

In this note, we restrict ourselves to considering solutions in the space of quadratic polynomials which give rise to linear Poisson brackets between invariant variables. Indeed, let us take quadratic potential 
\[
V_2(q_1,\ldots,q_n)=\sum_{i,j=1}^n V_{ij}q_iq_j\,,\qquad V_{ij}=V_{ji}\in \mathbb R\,,
\]

A pair of quadratic forms $V(q)=\sum q_i^2$ and $V_2(q)$ can simultaneously be reduced to a diagonal form by a linear orthogonal transformation. Thus, without loss of generality, we cold assume that 
\[Q_1=V(q)=\sum_{i=1}^n  u_iq_i^2\,,\qquad\mbox{and}\qquad Q_2=V_2(q)=\sum_{i=1}^n v_iq_i^2\,,\qquad u_i,v_i\in\mathbb R\,.\]  
In this note, we prefer not to change the variable $Q_1=r^2=q_1^2+\cdots+q_n^2$ and the corresponding kinetic energy
\[T=P_1=\sum_{i=1}^n p_i^2\,.\]
This allows us to compare our  invariant potential with the potentials published in \cite{f83,f86,d85,pe90,r86,rs94,rs03,tf94}.

Functions
\[Q_2=V_2(q_1,\ldots,q_n)\,,\qquad P_2=V_2(p_1,\ldots,p_n)\qquad\mbox{and}\qquad B=\{P_1,Q_2\}\]
are in the involution with $m$ functions $M_\alpha$, $\alpha=1,\ldots,m$, i.e. 
\[\{Q_2,M_\alpha\}=0\,,\quad\{P_2,M_\alpha\}=0\,,\quad \{B,M_\alpha\}=0\,,\qquad \alpha=1,\ldots,m\,,\]
Definition of $V_2(q)$ involves symmetric matrix $V_{ij}$ and, therefore, we have 
\[
B=\{P_1,Q_2\}=\{P_2,Q_1\}\,.
\]
The remaining unknown Poisson bracket
 \[
\{P_2,Q_2\}=\{V_2(p),V_2(q)\}=\left\{\sum_{i,j=1}^n V_{ij}p_ip_j,\sum_{k,\ell=1}^n V_{k\ell}q_kq_\ell \right\}
\]
is a linear function on $q$ and $p$ which is invariant with respect to rotations, i.e.
\[
\{P_2,Q_2\}=\alpha A+\beta B\,,\qquad \alpha,\beta\in \mathbb R\,.
\]
Changing the basis of solutions we can put
\bq\label{cond-V2}
A=\{P_1,Q_1\}=\{P_2,Q_2\}\,.
\eq
If condition (\ref{cond-V2}) holds, the Poisson brackets between six functions $Q_1,Q_2,P_1,P_2$ and $A,B$ are described by the following Poisson bivector
\bq\label{Pi}
\Pi=\left(
 \begin{array}{cccccc}
 0 & 0 & -A &-B & -8Q_1 & -8Q_2 \\
 0 & 0 & -B & -A& -8Q_2 & -8Q_1 \\
 A & B & 0 & 0 & 8P_1 & 8P_2 \\
 B & A & 0 & 0 & 8P_2 &8P_1 \\
 8Q_1 & 8Q_2 & -8P_1 &-8P_2 & 0 & 0 \\
 8Q_2 & 8Q_1 &-8P_2 & 8P_1 & 0 & 0 \\
 \end{array}
 \right)\,.
\eq
Two Casimir functions of this bivector $\Pi dC_i=0$ look like 
\bq\label{caz12}
C_1=A^2 + B^2 - 16Q_1P_1 - 16Q_2P_2\qquad\mbox{and}\qquad C_2=AB - 8Q_1P_2 - 8Q_2P_1\,.
\eq
They are polynomials of the second order on $m$ functions $M_\alpha$, $\alpha=1,\ldots,m$ (\ref{M-a}).

\begin{prop}
The Poisson bracket $\{.,.\}_\Pi$ defined by bivector $\Pi$ (\ref{Pi}) is reduced to the Lie-Poisson bracket
on the Lie algebra $so^*(4)$ over a complex number field. 
\end{prop}
Indeed, let us introduce variables 
\[
s_1=\frac{\mathrm i(A - B)}{16}\,,\qquad
s_2=-\mathrm i(P_1 - P_2) +\frac{\mathrm i(Q_1 - Q_2)}{64}\,,\qquad
s_3=-(P_1 - P_2) -\frac{Q_1-Q_2}{64}
\]
and
\[
t_1=\frac{\mathrm i(A + B)}{16}\,,\qquad
t_2=\mathrm{i}(P_1 + P_2) - \frac{\mathrm i(Q_1+Q_2)}{64}\,,\qquad
t_3=(P_1+P_2)+\frac{Q_1 + Q_2}{64} \,.
\]
The Poisson bracket $\{.,.\}_\Pi$ between these variables coincide with the standard Lie-Poisson brackets on $so^*(4)=so^*(3)\times so^*(3)$
\[
\{s_i,s_j\}=\epsilon_{ijk}s_k\,,\qquad \{t_i,t_j\}=\epsilon_{ijk}t_k\,,\qquad \{s_i,t_j\}=0\,,\qquad i,j,k=1,2,3.
\]
Here $\epsilon_{ijk}$ is a completely antisymmetric tensor.

Substitution of the variables $s_i$ and $t_i$ in integrals of motion for the Frahm-Shottky, Poincaré, Steklov and other well-known integrable systems on $so^*(4)$
we obtain integrable systems on $T^*\mathbb E^n$, which have no physical meaning.

\subsection{Integrable systems on $so^*(4)$}
In this section, we follow the calculations from the paper \cite{ts24}. 
Let us take invariant Hamiltonian $H$ (\ref{ham}) on $2n$-dimensional phase space $T^*\mathbb E^n$
 \bq\label{H-anz}
 H=\sum_{i=1}^n p_i^2+V(q_1,\ldots,q_n)=P_1+U(Q_1,Q_2)
 \eq
and calculate some families of integrable potentials $U(Q_1,Q_2)$ using a brute force method \cite{hiet87}. 
\begin{prop}
 Substituting $H$ (\ref{H-anz}) and the following polynomial of second order in momenta $p=(p_1,\ldots,p_n)$
 \bq\label{G-2}
 G=g_1P_1+g_2P_2+g_3A^2+g_4AB+g_5B^2+g_6\,, \qquad g_k\equiv g_k(Q_1,Q_2)
 \eq
in
\bq\label{poi-6}
\{H,G\}_\Pi=0
\eq
and solving the resulting equations in terms of $U(Q_1,Q_2)$ and $g_k(Q_1,Q_2)$ we obtain a family of integrable systems defined by
integrals of motion
\bq\label{H-2}
H=P_1+f_1(Q_1+Q_2)+f_2(Q_1-Q_2)\,,\qquad G=P_2+f_1(Q_1+Q_2)-f_2(Q_1-Q_2)\,.
\eq
 Here $f_1$ and $f_2$ are arbitrary functions.
\end{prop}
The proof is a straightforward solution of the equation (\ref{poi-6}), which allows us to prove that there are no other solutions than the solution (\ref{H-2}). 
This solution is related to the separation of variables in the Hamilton-Jacobi equations $H=E_1$ and $G=E_2$
\[
(P_1+P_2)+2f_1(Q_1+Q_2)=E_1+E_2\,,\qquad (P_1-P_2)+2f_2(Q_1-Q_2)=E_1-E_2\,.
\]

 To get something really interesting, we need to use a more complicated ansatz for the second integral of motion $G$.
\begin{prop}
 Substituting $H$ (\ref{H-anz}) and the following polynomial of fourth order in momenta $p=(p_1,\ldots,p_n)$
 \bq\label{G-4}
 G=P_2^2+g_1P_1^2+g_2 P_1P_2+ g_3P_1+g_4P_2+g_5A^2+g_6AB+g_7B^2+g_8\,, 
 \eq
 with $g_k\equiv g_k(Q_1,Q_2)$ in (\ref{poi-6}) and solving the resulting equations in terms of $U(Q_1,Q_2)$ and $g_k(Q_1,Q_2)$ we obtain two families of integrable systems with Hamiltonians
 \bq\label{H-first}
 H_{I}=P_1+a_1 (2Q_1^2 - Q_2^2) + a_2Q_1+\frac{a_3Q_1}{Q_1^2 - Q_2^2} + \frac{a_4Q_2}{Q_1^2 - Q_2^2} \,,\qquad a_i\in\mathbb R
 \eq
and
\bq\label{H-sec}
H_{II}=P_1+a_1(5Q_1 + 3Q_2)(3Q_1 + Q_2)+a_2(5Q_1 + 3Q_2)+ \frac{a_3}{Q_1 + Q_2} +\frac{a_4}{Q_1 - Q_2} + \frac{a_5}{(Q_1 - Q_2)^3} \,.
\eq
\end{prop}
The first part of the proof is a straightforward solution of (\ref{poi-6}). In the second part, we prove that there are no 
 polynomials $G$ of second order in momenta commuting with these Hamiltonians.

Let us explicitly present integrals of motion $G$ (\ref{G-4}) for the Hamiltonians $H_{I}$ (\ref{H-first}) and $H_{II}$ (\ref{H-sec}) 
\begin{align*}
G_{I}&= P_2^2 + 2a_1Q_2^2P_1- 2\left(a_2Q_2 + \frac{a_4Q_1 +a_3Q_2}{Q_1^2 - Q_2^2}\right)P_2-\frac{a_1Q_2AB}{2} +\frac{ (2a_1Q_1 + a_2)B^2}{4} 
\\
&+Q_2^2(a_1^2Q_2^2 + 2a_1a_2Q_1 + a_2^2) + \frac{2a_1Q_2^2(a_3Q_1 +a_4 Q_2)}{Q_1^2 - Q_2^2}
 + \frac{2a_2Q_2\left(a_4Q_1 +a_3 Q_2\right)}{Q_1^2 - Q_2^2} 
 \\
 &+\frac{ (a_4Q_1 +a_3 Q_2)^2}{(Q_1^2 - Q_2^2)^2}
\end{align*}
and
\begin{align*}
&G_{II}= P_2^2 -2P_1P_2 -4\left((Q_1 - Q_2)\bigl(2a_1(Q_1 + Q_2) + a_2\bigr) + \frac{a_5}{(Q_1 - Q_2)^3} +\frac{a_4}{Q_1 - Q_2}\right)P_2 
\\
&-2\left(a_1(Q_1 + 3Q_2)(7Q_1 + Q_2) + a_2(3Q_1 + 5Q_2) + \frac{a_3}{Q_1 + Q_2} -\frac{a_4}{Q_1 - Q_2}-\frac{a_5}{(Q_1 - Q_2)^3}\right)P_1
\\
 &+2a_1(A - B)(AQ_2 - BQ_1)
 -2 a_1 (11 a_3 + 3 a_4) Q_1 - 2 a_1 Q_2 (11 a_3 + 25 a_4) 
 \\
 &- (3 Q_1 + 5 Q_2) (7 Q_1 + Q_2) (3 Q_1 a_1 + Q_2 a_1 + a_2)^2 -\frac{a_3^2}{(Q_1 + Q_2)^2} + \frac{3 a_4^2}{(Q_1 - Q_2)^2} + \frac{3 a_5^2}{(Q_1 - Q_2)^6}
 \\
 &-\frac{2 a_3 a_4}{Q_1^2 - Q_2^2} - \frac{2 a_3 a_5}{(Q_1 + Q_2) (Q_1 - Q_2)^3} +\frac{ 6 a_4 a_5}{(Q_1 - Q_2)^4} + \frac{4a_3 (6a_1 Q_2 + a_2) Q_2}{Q_1 + Q_2}
 \\
 &- \frac{16a_2 Q_2 (4a_1 Q_2 + a_2)}{Q_1 - Q_2} - \frac{16a_5 (2a_1 Q_2^2 -a_2 Q_2)}{(Q_1 - Q_2)^3} + \frac{2 a_5 (4a_1 Q_2 - a_2)}{(Q_1 - Q_2)^2}
 +\frac{ 26 a_1 a_5}{Q_1 - Q_2}\,.
 \end{align*}

One more partial solution was obtained in \cite{d86,ts15,ts15a} and then generalized in \cite{ts22a,ts22b}.
\begin{prop}
Let us take Hamiltonian $H$ (\ref{H-anz}) and the following polynomial of fourth order in momenta $p=(p_1,\ldots,p_n)$
 \begin{align}
 G&=(Q_1 - Q_2)(P_2^2+b_1P_1P_2) + (b_2A^2 +b_3 AB +b_4 B^2)P_2 + (b_5A^2 + b_6AB +b_7 B^2)P_1
 \nonumber\\
 &+g_1P_1 +g_2P_2 + g_3A^2 + g_4AB +g_5B^2 + g_6\,, \label{G-43}
 \end{align}
 depending on parameters $ b_k\in\mathbb R$ and functions $ g_k(Q_1,Q_2)$. Solving equation (\ref{poi-6}) 
 we obtain an integrable system with the following Hamiltonian
 \begin{align}
 H_{III}=P_1+ &a_1(29Q_1^2 - 30Q_1Q_2 + 5Q_2^2) + a_2(5Q_1 - 3Q_2)\label{H-third}
\\ 
 +&\frac{a_3}{Q_1 - Q_2} + \frac{a_4}{Q_1 + Q_2} +\frac{a_5(5Q_1 - 3Q_2)}{(Q_1 + Q_2)^3}
 \,,\qquad a_i\in\mathbb R
 \nonumber
 \end{align}
\end{prop}
As above we can prove that equation (\ref{poi-6}) has this solution and that there are no 
quadratic polynomials $G$ (\ref{G-2}) commuting with Hamiltonian $H_{III}$ (\ref{H-third}).

For the sake of completeness, we explicitly present values of parameters
\[
b_1=3\,,\quad b_2= -3/16\,,\quad b_3= -1/8\,,\quad b_4=1/16\,, \quad b_5=b_6=0\,,\quad b_7= 1/4\,,
\]
and functions
\begin{align*}
g_1=&
a_1(Q_2 - Q_1)(45Q_1^2 - 30Q_1Q_2 - 11Q_2^2) -a_2 (9Q_1 + Q_2)(Q_1 - Q_2) + \frac{6a_4Q_1}{Q_1 + Q_2}
\\
 +& \frac{3a_5(9Q_1 + Q_2)(Q_1 - Q_2)}{(Q_1 + Q_2)^3}\,,
 \\
 g_2=&a_1(Q_2 - Q_1) (3 Q_1^2 + 62 Q_1 Q_2 - 5 Q_2^2) +a_2 (Q_1 - Q_2) (Q_1 - 7 Q_2) + a_3
+\frac{ 5a_4 (Q_1 - Q_2)}{Q_1 + Q_2}
 \\
 +&\frac{a_5 (21 Q_1 - 19 Q_2) (Q_1 - Q_2) }{(Q_1 + Q_2)^3}\,,
 \\
16 g_3=&a_1(45 Q_1^2 - 30 Q_1 Q_2 - 11 Q_2^2) +a_2 (9 Q_1 + Q_2) + \frac{3 a_3}{Q_1 - Q_2} -\frac{ 3 a_4}{Q_1 + Q_2}
\\
 -&\frac{ 3a_5 (9 Q_1 + Q_2) }{(Q_1 + Q_2)^3}\,,
 \\
 8g_4=&a_1(-21Q_1^2 + 46Q_1Q_2 + 3Q_2^2) +a_2 (3Q_2 - 5Q_1) +\frac{a_3}{Q_1 - Q_2} -\frac{a_4}{Q_1 + Q_2}
 \\
 +&\frac{a_5 (3Q_1 + 11Q_2)}{(Q_1 + Q_2)^3}\,,
 \\
 16g_5=&a_1
 (-3Q_1^2 - 62Q_1Q_2 + 5Q_2^2) +a_2 (Q_1 - 7Q_2) +\frac{ 3a_3}{Q_1 - Q_2} +\frac{ 5a_4}{Q_1 + Q_2}
 \\
 +&\frac{ a_5(21Q_1 - 19Q_2)}{(Q_1 + Q_2)^3}\,,
\end{align*}
and
\begin{align*}
g_6=&4(Q_1 - Q_2)(Q_1 + Q_2)^2(6a_1Q_1 - 2a_1Q_2 + a2)^2+a_2\Bigl(a_3(5Q_1 - 3Q_2) +a_4 (7Q_1 - Q_2)\Bigr)
\\
+&a_1\Bigl(
a_3 (45Q_1^2 + 2Q_1Q_2 + 21Q_2^2)
+a_4(39Q_1^2 - 26Q_1Q_2 - Q_2^2) - 32a_5 (9Q_1 - Q_2) \Bigr)
\\+&\frac{a_3(a_3 + 3a_4)}{Q_1 - Q_2}+\frac{1024a_1a_5Q_1^2 + 96a_2a_5Q_1 + a_3a_4 - a_4^2}{Q_1 + Q_2}
\\
-&\frac{1024a_1a_5Q_1^3 + 128a_2a_5Q_1^2 - 8a_4^2Q_1 - 21a_3a_5 - 63a_4a_5)}{(Q_1 + Q_2)^2}
-\frac{8a_5(a_3+ 23a_4)Q_2}{(Q_1 + Q_2)^3}\\
 +& \frac{128a_4a_5Q_2^2}{(Q_1 + Q_2)^4 }+\frac{16a_5^2(Q_1 - Q_2)(3Q_1-Q_2)^2}{(Q_1 + Q_2)^6}\,.
\end{align*}
It allows directly verify that $H_{III}$ (\ref{H-third}) commutes with $G_{III}$ (\ref{G-43}).

Since the solutions of the equations $\mathcal L_{Y_a}V=0$ (\ref{m-eq}) and (\ref{cond-V2}) are defined up to the sign, all the results are valid after the transformation of variables
\bq\label{trans}
Q_2\to -Q_2\,,\qquad P_2\to -P_2\,,\qquad B\to-B\,.
\eq
The first Hamiltonian $H_{I}$ (\ref{H-first}) is invariant to this transformation (\ref{trans}). Second and third Hamiltonians (\ref{H-sec},\ref{H-third})
are changed after the transformation (\ref{trans}) and we get two more families of integrable systems
 \bq\label{H-4}
H_{IV}=P_1+a_1(5Q_1 - 3Q_2)(3Q_1 - Q_2)+a_2(5Q_1 - 3Q_2)
 +\frac{a_3}{Q_1 - Q_2} +\frac{a_4}{Q_1+ Q_2} + \frac{a_5}{(Q_1 + Q_2)^3} 
\eq
and 
\bq
 H_{V}=P_1+a_1(29Q_1^2 + 30Q_1Q_2 + 5Q_2^2) + a_2(5Q_1 + 3Q_2)
 +\frac{a_3}{Q_1 + Q_2} + \frac{a_4}{Q_1 - Q_2} +\frac{a_5(5Q_1 + 3Q_2)}{(Q_1 - Q_2)^3}\,.\label{H-5}
 \eq
 Thus we have five families of integrable systems associated with arbitrary pair of independent solutions of the equations (\ref{m-eq}) and (\ref{cond-V2}) 
 in $n$-dimensional Euclidean space $\mathbb E^n$.

If we put $a_2=a_3=a_4=a_5=0$, then these Hamiltonians have the following form
\[
H_N=\sum_{i=1}^n p_i^2+
\sum_{\alpha,\beta,\gamma,\delta} \mathcal R_{-\alpha,\beta,\gamma,-\delta}q^\alpha q^\beta q^\gamma q^\delta\,.
\]
The fourth-order tensors $\mathcal R$ can sometimes be identified with the curvature tensor or Riemann tensor of manifolds not directly related to Euclidean space. Examples of integrable systems for which $\mathcal R$ coincide with the constant curvature tensor of the Hermitian symmetric spaces can be found in \cite{f83,d86,f86,ts23,ts23a}.
Here we not only reproduce these results in the framework of the symplectic reduction theory but also add new rational terms to these potentials. 

 Below we will present a few examples of the integrable Hamiltonians and, for brevity, we will omit explicit formulae for integrals of motion $G_{N}$, $N=I,II,III,IV,V$.

\section{Abelian subgroups of $SO(n)$ at $n\leq 4$}
Each plane through the rotation centre  is the axis-plane of a commutative subgroup isomorphic to $SO(2)$. All these subgroups are mutually conjugate in $SO(n)$. Each pair of completely orthogonal planes through the rotation centre  is the pair of invariant planes of a commutative subgroup of $SO(4)$ isomorphic to $SO(2) \times SO(2)$.  

So, the $m=n-2$ commuting rotations of Euclidean space $\mathbb E^n$ exists only at $n\leq 4$ \cite{cayley}. The corresponding symmetry fields $Y_\alpha$, $\alpha=1,\ldots,m$, commute with each other and generate linear conservation laws $M_\alpha$ (\ref{M-a}) in the involution with respect to the canonical Poisson bracket (\ref{poi}).

\subsection{Three dimensional Euclidean space $\mathbb E^3$}
Let us consider the so-called reduced representation of the abelian subgroup $SO(2)$ of $SO(3)$. The corresponding conservation law is equal to
\[
M_1=J_{12}=q_1p_2-q_2p_1\,.
\]
Equation (\ref{m-eq}) has a linear solution $V(q)=q_3$, that allows us to get two independent solutions in the space of polynomials of second-order
\[
V_1=q_1^2+q_2^2+q_3^2\quad\mbox{and}\quad V_2=q_1^2+q_2^2 - q_3^2\,,
\]
for which condition (\ref{cond-V2}) holds. Substituting variables
\bq\label{rel-3}
Q_{1,2}=V_{1,2}(q)\,,\qquad P_{1,2}=V_{1,2}(p)\,,\quad A=\{P_1,Q_1\}\,,\quad B=\{P_1,Q_2\}
\eq
into (\ref{H-first}), (\ref{H-sec}) and (\ref{H-third}) we obtain the following integrable Hamiltonians
\begin{align*}
H_{I}=\sum_{i=1}^3 p_i^2+&a_1(q_1^4 + 2q_1^2q_2^2 + 6q_1^2q_3^2 + q_2^4 + 6q_2^2q_3^2 + q_3^4) + a_2(q_1^2 + q_2^2 + q_3^2)
\\
+&
\frac{a_3-a_4}{4(q_1^2 + q_2^2)} +\frac{a_3+a_4}{4q_3^2}\,,
\\
H_{II}=\sum_{i=1}^3 p_i^2+&4a_1(4q_1^2 + 4q_2^2 + q_3^2)(2q_1^2 + 2q_2^2 + q_3^2) + 2a_2(4q_1^2 + 4q_2^2 +q_3^2)
\\
+&\frac{a_3}{2(q_1^2 + q_2^2)} + \frac{a_4}{2q_3^2} + \frac{a_5}{8q_3^6}\,,
\\
H_{III}=\sum_{i=1}^3 p_i^2+&
4a_1\left(q_1^4 + 2q_1^2q_2^2 + 12q_1^2q_3^2 + q_2^4 + 12q_2^2q_3^2 + 16q_3^4\right) +a_2\left(q_1^2 + q_2^2 + 4q_3^2\right)
\\
 +&\frac{a_3}{2q_3^2} + \frac{a_4}{2(q_1^2 + q_2^2)} +\frac{ a_5(q_1^2 + q_2^2 + 4q_3^2)}{4(q_1^2 + q_2^2)^3}\,.
\end{align*}
Two remaining Hamiltonians (\ref{H-4}) and (\ref{H-5}) are equal to
\begin{align*}
H_{IV}=\sum_{i=1}^3 p_i^2+&4a_1(q_1^4 + 2 q_1^2 q_2^2 + 6 q_1^2 q_3^2 + q_2^4 + 6 q_2^2 q_3^2 + 8 q_3^4)
+a_2 (2 q_1^2 + 2 q_2^2 + 8 q_3^2) 
\\+&\frac{ a_3}{2 q_3^2} +\frac{ a_4}{2(q_1^2 + q_2^2)} + \frac{a_5}{(2q_1^2 + 2 q_2^2)^3}\\
\end{align*}
and
\begin{align}
{H}_{V}=\sum_{i=1}^3 p_i^2+&
4a_1\left(16q_1^4 + 32q_1^2q_2^2 + 12q_1^2q_3^2 + 16q_2^4 + 12q_2^2q_3^2 + q_3^4\right) + 2a_2\left(4q_1^2 + 4q_2^2 + q_3^2\right)
\nonumber
\\
 +& \frac{a_3}{2(q_1^2 + q_2^2)} + \frac{a_4}{2q_3^2} + \frac{a_5(4q_1^2 + 4q_2^2 + q_3^2)}{4q_3^6}\,.\label{H-sup3}
\end{align}
The Casimir functions $C_{1,2}$ (\ref{caz12}) for this realization (\ref{rel-3}) of algebra $so^*(4)$ are equal to 
\[
C_1=-32M_1^2\,,\qquad C_2=-16M_1^2\,.
\]
Since the Hamiltonians $H_{N}$, $N=I,II,III,IV,V$ commute with $M_1$ and $G_{K}$
\[
\{H_{N},M_1\}=0\,,\qquad \{H_{N},G_{N}\}=0\,,\quad \{G_{N},M_1\}=0\,,
\]
we have five integrable systems with three degrees of freedom. The polynomial terms in these five potentials were obtained in \cite{d86}. 

 Rational terms in these Hamiltonians also occurred in the literature, for example, the Hamiltonian $H_{V}$ at 
 $a_1=a_2=a_3=a_4=0$ was found in \cite{ts15,ts15a} as an example of a non-trivial superintegrable system. 
 Indeed, the Hamiltonian $H$ (\ref{H-sup3}) at $a_1=a_2=a_3=a_4=0$ 
\[
{H}_{V}=p_1^2+p_2^2+p_3^2+ \frac{a_5(4q_1^2 + 4q_2^2 + q_3^2)}{4q_3^6}
\]
is invariant to the symmetry field $X_{12}$ and commutes with the fourth degree polynomial $G_{V}$. According to \cite{ts22a, ts22b} 
this Hamiltonian also commutes with two polynomials of second order in momenta 
\begin{align}
K_1=&p_3J_{23} - 2p_1J_{12} + \frac{2a_5q_2(2q_1^2 + 2q_2^2 + q_3^2)}{q_3^6}\, \nonumber\\
\label{k-super}\\
K_2=&p_3J_{13} + 2p_2J_{12} + \frac{2a_5q_1(2q_1^2 + 2q_2^2 + q_3^2)}{q_3^6}\,,\nonumber
\end{align}
which are not invariant to the symmetry field $X_{12}$. The algebra of first integrals has the form
\[
\{M_1,K_1\}=-K_2\,,\quad \{M_1,K_2\}=K_1\,,\quad \{K_1,K_2\}=4M_1 {H}_{V}\,,\quad M_1=J_{12}\,,
\]
and the polynomial of the fourth-degree $G_V$ is an element of the centre of this algebra
\[
G_{V}=4M_1^2{H}_{V} - K_1^2 - K_2^2\,.
\]
Since functions $K_{1,2}$ are not invariant with respect to rotation $X_{12}$ we can not find these integrals of motion within our scheme. 
A generalization of this superintegrable system to the $n$-dimensional case may be found in \cite{ts22a,ts22b}.

Thus we have completely reproduced the list of fourth-degree polynomial potentials given in Table 1 of \cite{d86}, except for the first two cases in that table, where the fourth-degree polynomial $G$ is a square of the second-degree polynomial on momenta. 

To get the corresponding Fordy-Kulish system we have to take double rotation $Y_1=X_{12}+X_{23}$ and the corresponding conservation law
\[
M_1=J_{12}+J_{23}=(q_1p_2-p_1q_2) + (q_2p_3 - p_2q_3)\,.
\] 
Of course, using rotation $X_{13}$ we can reduce double rotation $Y_1$ to a single rotation $X_{23}$ and obtain all the previous Hamiltonians up to the permutation of indexes. 

Nevertheless, allow us to reiterate one of these results. Equations (\ref{m-eq}) and (\ref{cond-V2}) have two independent solutions 
\[ V_1=q_1^2+q_2^2+q_3^2\quad\mbox{and}\quad V_2=2q_1q_3 - q_2^2\,\]
in the class of polynomials of second order. In this case first integrable Hamiltonian (\ref{H-first}) has the form
\begin{align*} 
H_{I}=\sum_{i=1}^3 p_i^2+&a_1\left(2(q_1^2 + q_2^2 + q_3^2)^2 - (2q_1q_3 - q_2^2)^2\right)+a_2(q_1^2 + q_2^2 + q_3^2)
\\
+&\frac{a_3(q_1^2 + q_2^2 + q_3^2)+a_4(2q_1q_3 - q_2^2)}{(q_1 + q_3)^2(q_1^2 - 2q_1q_3 + 2q_2^2 + q_3^2)}\,.
\nonumber
 \end{align*}
 When $a_2=a_3=a_4=0$ this Hamiltonian is associated with the Hermitian symmetric space $Sp(3)/U(3)$ of $\mathrm{C.I}$ type, see \cite{f83,f86}. We know Lax matrices and the corresponding classical $r$-matrix only in this partial case. It will be interesting to construct Lax matrices and the corresponding classical $r$-matrix for other integrable Hamiltonians discussed in this section.

\subsection{Four-dimensional Euclidean space $\mathbb E^4$}
Let us start by constructing integrable systems invariant with respect to two commuting basis fields of symmetries $X_{1,2}$ and $ X_{34}$.
In this case equations (\ref{m-eq}) and (\ref{cond-V2}) have two independent solutions 
\[
Q_1=V_1(q)=q_1^2+q_2^2+q_3^2+q_4^2\,,\qquad Q_2=V_2(q)=q_1^2+q_2^2-q_3^2-q_4^2\,.
\]
Substituting $Q_2=\pm V_2(q)$ into $H_{N}$, $N=I,II,III,IV,V$, we obtain explicit expressions for five integrable Hamiltonians in Euclidean space $\mathbb E^4$ associated with the reduced representation of abelian subgroup $SO(2)\times SO(2)\subset SO(4)$.

\subsubsection{Double rotations}
Let us consider non-reduced representations of the same abelian subgroup $SO(2)\times SO(2)\subset SO(4)$ associated with 
the following symmetry fields 
\[Y_{1}=X_{12}+X_{34}\,,\qquad Y_{2}=X_{13}+X_{24}\,.\]
The corresponding linear polynomials in momenta
 \bq\label{rot4n}
 M_1= J_{1 2} + J_{3 4}\qquad\mbox{and}\qquad M_2=J_{13 }+ J_{24}\,,\qquad \{M_1,M_2\}=0\,,
 \eq
are in the involution for each other. Equations (\ref{m-eq}) and (\ref{cond-V2})
 have two independent solutions 
\[
V_1=\sum_{i=1}^4 q_i^2\qquad\mbox{and}\qquad V_2=2(q_1q_4 - q_2q_3)\,.
\] 
In partial case $a_2=a_3=a_4=0$ first Hamiltonian (\ref{H-first}) look like
\[
H_{I}=\sum_{i=1}^4 p_i^2+a_1\left(2\left(\sum_{i=1}^4 q_i^2 \right)^2- (q_1q_4 - q_2q_3)^2\right)\,.
 \]
This Hamiltonian is associated with the curvature tensor on the Hermitian symmetric space $SO(4)/S\left(U(2)\times U(2)\right)$ of $\mathrm{A.III}$ type \cite{f86,ts23,ts23a}.

When $a_2=a_3=a_4=a_5=0$ second and third Hamiltonians (\ref{H-sec},\ref{H-third}) are equal to 
\begin{align*}
&H_{II}=P_1+a_1(5Q_1 + 3Q_2)(3Q_1 + Q_2)\\
=&\sum_{i=1}^4 p_i^2+
a_1(5q_1^2 + 6q_1q_4 + 5q_2^2 - 6q_2q_3 + 5q_3^2 + 5q_4^2)(3q_1^2 + 2q_1q_4 + 3q_2^2 - 2q_2q_3 + 3q_3^2 + 3q_4^2)
\end{align*}
and
\begin{align*}
H_{III}=P_1+ &a_1(29Q_1^2 - 30Q_1Q_2 + 5Q_2^2)\\
\sum_{i=1}^4 p_i^2+&a_1\left(29 q_1^4 - 60 q_1^3 q_4 + 58 q_1^2 q_2^2 + 60 q_1^2 q_2 q_3 + 58 q_1^2 q_3^2 + 78 q_1^2 q_4^2 - 60 q_1 q_2^2 q_4\right.
\\
& - 40 q_1 q_2 q_3 q_4 - 60 q_1 q_3^2 q_4 - 60 q_1 q_4^3 + 29 q_2^4 + 60 q_2^3 q_3 + 78 q_2^2 q_3^2 + 58 q_2^2 q_4^2
\\ \\
&\left. + 60 q_2 q_3^3 + 60 q_2 q_3 q_4^2 + 29 q_3^4 + 58 q_3^2 q_4^2 + 29 q_4^4\right)\,.
\end{align*}
These two Hamiltonians are absent in the list of integrable systems with quartic potentials associated with the Hermitian symmetric spaces \cite{f86}.
Other Hamiltonians $H_{IV}$ and $H_V$ are also new.

\subsubsection{Triple rotations}
Two linear polynomials associated with double and triple rotations in Euclidean space $\mathbb E^4$
 \bq\label{rot-32}
 M_1= J_{1 2} + J_{3 4}\qquad\mbox{and}\qquad M_2=J_{1 2} + J_{1 3} + J_{2 4}\,,\qquad \{M_1,M_2\}=0
 \eq
 commute for each other. In this case equations (\ref{m-eq}) and (\ref{cond-V2}) have two independent solutions
\[
V_{1}=\sum_{i=1}^4 q_i^2\qquad\mbox{and}\qquad V_2=\frac{\sqrt{5}}{5} \left(q_1^2 - 4q_1q_4 + q_2^2 + 4q_2q_3 - q_3^2 - q_4^2\right)\,.
\] 
The corresponding first Hamiltonian (\ref{H-first}) is
 \[
 H_{I}=\sum_{i=1}^4 p_i^2+a_1\left(2 \left(\sum_{i=1}^4 q_i^2\right)^2-\frac{1}{5} \left(q_1^2 - 4q_1q_4 + q_2^2 + 4q_2q_3 - q_3^2 - q_4^2\right)^2\right)\,.
 \]
 Since the triple rotation $Y_2=X_{1 2} + X_{1 3} + X_{2 4}$ consists of principal rotations $X_{ij}$ with the coinciding indices, it could be reduced to other forms using Cayley transforms \cite{cayley}.  Discussion of the corresponding reductions of invariant potentials to some canonical form is beyond the scope of this note devoted to the algorithm of construction of invariant potentials.
 
\section{Noncommutative subgroups of $SO(n)$ at $n>4$}
For $m>n-2$, rotation symmetry fields $Y_\alpha$, $\alpha=1,\ldots,m$ may not commute with each other, but we will nevertheless require the existence of $n-2$ linear or quadratic conservation laws in involution concerning Poisson brackets (\ref{poi}).

In fact at $m>n-2$ we inevitably obtain so-called superintegrable systems which possess more globally defined conserved quantities than their degrees of freedom \cite{bbm14,mil,resh}. They are also known as non-commutative integrable systems and as systems with degenerate integrability. Locally these integrals of motion are functions on the action variables $I_j$ and also on the angle variables $\varphi_i$ which are usually expressed via Abels' quadratures \cite{ts08}.  It allows us to get global integrals of motion by using the Euler theorem \cite{ts08a,ts09}, Chaplygin theorem \cite{ts18}, Riemann-Roch theorem \cite{ts20}, and other classical theorems appearing in various mathematical applications \cite{ts19r,ts19s}.

In the framework of the symplectic reduction theory we can not construct additional global integrals of motion at $m=n-2$, since these integrals of motion will not be invariant to our chosen symmetry fields, see example (\ref{k-super}). If $m>n-2$ we inevitably obtain additional integrals of motion for superintegrable systems, since the symmetry fields do not commute with each other.

Further, we restrict ourselves to two examples from the set of degenerate by Kolmogorov or superintegrable systems with integrals of motion which are polynomial of higher order in momenta. Integrals of motion which are polynomials of second order in momenta are discussed in the review \cite{mil}.

\subsection{Five-dimensional Euclidean space $\mathbb E^5$}
Let us consider left and right isoclinic double rotations 
\[Y_{1}=X_{12}+X_{34}\,,\qquad Y_{2}=X_{13}+X_{24}\]
which we used to construct integrable systems in four-dimensional Euclidean space $\mathbb E^4$ and 
add two linear polynomials 
\[M_3=J_{25} - J_{3 5}\,,\qquad M_4= J_{15} + J_{45}\]
to a pair of commuting linear polynomials 
\[
 M_1= J_{1 2} + J_{3 4}\qquad\mbox{and}\qquad M_2=J_{13}+ J_{24}\,,\qquad \{M_1,M_2\}=0\,,
 \]
 so that
 \[
 \{M_1, M_3\}=- M_4\,,\quad \{M_1, M_4\}= M_3\,, \quad
\{M_2, M_3\}=M_4\,,\quad \{M_2 M_4\} =- M_3\,.
 \]
 It allows us to construct $n-2=3$ independent functions in the involution
 \[
 \{M_1,M_3^2+M_4^2\}=0\,,\qquad \{M_2,M_3^2+M_4^2\}=0\,.
 \]
 associated with $m=4$ functions $M_1,M_2,M_3,M_4$.
 
 Invariant to the chosen four symmetry fields $Y_\alpha$ potentials are equal to 
 \[
 V_1(q)=\sum_{i=1}^n q_i^2\quad \mbox{and}\quad V_2(q)=\pm(2q_1q_4 - 2q_2q_3 + q_5^2)\,.
 \]
 The corresponding variables 
 \[
Q_{1,2}=V_{1,2}(q)\,,\qquad P_{1,2}=V_{1,2}(p)\,,\quad A=\{P_1,Q_1\}\,,\quad B=\{P_1,Q_2\}
\]
 satisfy to condition (\ref{cond-V2}). 
 
Thus, we obtain five superintegrable systems on the phase space $T^*\mathbb E^5$ with integrals of motion $M_1,M_2,M_3,M_4$ and $H_N,G_N$, $N=I,II,III,IV,V$. Two partial cases of these Hamiltonian systems are related to the Hermitian symmetric spaces of B.III and BD.I type, see \cite{ts23,ts23a}.
 
 \subsubsection{Unreduced composition of rotations}
 Let us add two polynomials
 \[
 M_3=(5\sqrt{5}-11)J_{1 5} + (3 - \sqrt{5})J_{25} + (1 - \sqrt{5})J_{3 5} + (7 - 3\sqrt{5})J_{45}
 \]
 and
 \[
 M_4=(3 - \sqrt{5})J_{15} + (11 - 5\sqrt{5})J_{25} + (7 - 3\sqrt{5})J_{35} - (1 - \sqrt{5})J_{45}
 \]
 to polynomials (\ref{rot-32}) associated with the double and triple rotations
 \[
 M_1= J_{1 2} + J_{3 4}\qquad\mbox{and}\qquad M_2=J_{1 2} + J_{1 3} + J_{2 4}\,,\,.
 \]
Because 
\[
\{M_1,M_2\}=0\,,\qquad \{M_1,M_3^2+M_4^2\}=0\,,\qquad \{M_2,M_3^2+M_4^2\}=0
\]
we have three commuting integrals of motion in $T^*\mathbb E^5$ in addition to $H_N$ and $G_N$. Solving equations (\ref{m-eq}) and (\ref{cond-V2}) we find invariant variables
\[Q_1=\sum_{i=1}^5 q_i^2\,,\qquad
Q_2=\pm\frac{\sqrt{5}\left( q_1^2 - 4q_1q_4 + q_2^2 + 4q_2q_3 - q_3^2 - q_4^2-\sqrt{5}q_5^2\right)}{5}\,.
\]
Substituting these variables into $H_N$, $N=I,II,III,IV,V$, we formally obtain five superintegrable Hamiltonians, which are practically unobservable in the original coordinates $q_1,\ldots,q_5$ without simplification. 

Thus, by solving equations (\ref{m-eq}) and (\ref{cond-V2}) for a given set of symmetry fields $Y_\alpha$, it is easy to construct integrable and superintegrable systems, which require further investigation. One of the main problems is how to reduce invariant potentials 
$V_N$ to some canonical form, since we can reduce potentials either to the shortest form or to the form associated with a hermitian symmetric space. 

In symplectic reduction theory, we do not have this problem, because the canonical form is an expression in the invariant variables $Q_1$ and $Q_2$ instead of various possible expressions in Cartesian coordinates $q_1,\ldots,q_n$.

\section{Conclusion}
To construct integrable and superintegrable systems in $n$-dimensional Euclidean space we propose to use invariant to $m$ rotational isometries $Y_\alpha$ variables, which are constructed from the kinetic and potential parts of the Hamiltonian 
\[P_1=T=\sum_{i=1}^2 p_i^2\,, \qquad Q_2=V(q)\]
where $V(q)$ is the solution of equations
\[
 \mathcal L_{Y_\alpha}V(q)=0\,,
 \] 
in the space of the second-order polynomials in Cartesian coordinates $q=(q_1,\ldots,q_n)$. 

The first natural generalisation of this algorithm is to take solutions of the same equations in the space of polynomials of higher order, which leads to the nonlinear Poisson brackets between invariant variables. For instance, substituting the following realization of the subgroup $SO(3)\times SO(2)\subset SO(6)$
\[
Y_1=X_{12}+X_{45}\,,\quad Y_2=X_{13}+X_{46}\,,\qquad Y_3=X_{23}+X_{56}
\]
and
\[Y_4=X_{14}+X_{25} + X_{36}\]
into the equations (\ref{m-eq}) and solving the resulting system of equations we obtain the standard solution in the space of polynomials of second order $V_1(q)=\sum q_i^2$ and one more complicated solution 
\[V_2(q)=2q_2q_5(q)_1q_4 + q_3q_6)-(q_1^2 - q_3^2)q_5^2-(q_4^2 - q_6^2)q_2^2 - (q_1q_6 - q_3q_4)^2
\] 
in the space of polynomials of fourth order associated with the symmetric space $SO(6)/S(U(3)\times U(2))$, see \cite{f83}. These invariant variables form polynomial algebra which can be used to construct superintegrable systems in $\mathbb E^6$.

The second natural generalisation is to completely or partially replace isometries with hidden symmetries, i.e. replace Killing vectors on the Killing tensors.
For instance, the generic Killing tensor of valency two in Euclidean space is given by
 \[
 K=\sum_{i,j} a_{ij} X_i\circ X_j+\sum_{i,j,k} b_{ijk} X_i\circ X_{j,k}+\sum_{i,j,k,m} c_{ijkm} X_{i,j}\circ X_{k,m}\,,
 \]
 where $X_i$ and $X_{ij}$ are basic isometries
\[
 X_i=\partial _i \qquad X_{i,j}=q_iX_j-q_jX_i\,,\qquad \partial_k=\frac{\partial}{\partial q_k}\,,
 \] 
 $a_{ij}, b_{ijk}$ and $c_{ijm}$ are parameters and $\circ$ denotes symmetric product. 
 We can take $m$ Killing tensors $K_\alpha$ and try to get invariant variables $Q_i=V_{i}(q_1,\ldots,q_n)$ solving equations
 \[
 d\left(K_\alpha dV\right)=0
 \]
instead of equations 
\[\mathcal L_{Y_\alpha} V=0\,, \] 
where $Y_\alpha$ are $m$ Killing vectors. Examples of such Killing tensors $K_\alpha$ with nonzero Haantjes torsion and the corresponding integrable systems can be found in \cite{ts23,ts23a}. We plan to reproduce these results in the framework of the symplectic reduction theory, which allows us also to get new integrable and superintegrable systems in Euclidean space.
 
It will be interesting also to apply the proposed algorithm to other metric spaces with a fairly large isometry group.
 
This work was supported by the Russian Science Foundation under grant no. 19-71-30012, https://www.rscf.ru/project/23-71-33002/.

\end{document}